# Non-Hamiltonian Holes in Grid Graphs

Heping, Jiang

Rm. 702, Building 12, Luomashi Avenue, XuanWu district,

Beijing, China, zip code: 100052

E-mail: jjhpjhp@gmail.com

## Abstract

In this paper we extend general grid graphs to the grid graphs consist of polygons tiling on a plane, named polygonal grid graphs. With a cycle basis satisfied polygons tiling, we study the cyclic structure of Hamilton graphs. A Hamilton cycle can be expressed as a symmetric difference of a subset of cycles in the basis. From the combinatorial relations of vertices in the subset of cycles in the basis, we deduce the formula of inside faces in Grinberg theorem, called Grinberg equation, and derive a kind of cycles whose existence make a polygonal grid graph non-Hamiltonian, called non-Hamiltonian holes, and then we characterize the existence condition of non-Hamiltonian holes and obtain the necessary and sufficient condition of a polygonal grid graph to be Hamiltonian. The result in this paper provides a new starting point for developing a polynomial-time algorithm for Hamilton problem in general grid graphs.

## 1. Introduction

In this paper we consider finite, undirected, connected, and simple (no loops or parallel edges) graphs only. Terms and notations not defined here can be found in **[1]**.

A grid graph is a subset of the integer lattice consisting of the tiling of the plane with unit squares. A Hamilton graph is a graph contains a Hamilton cycle (a cycle containing every vertex of the graph). Hamilton problem is to determine conditions under which a graph contains such cycle. 1982, Itai et al. show that the Hamilton problem for general grid graphs remains NP-complete **[2]**. 1995, Hwan-Gue, Cho and A. Zelikovsky conjecture that the Hamilton problem for solid grid graphs is NP-complete **[3]**. 1997, C. Umans and W. Lenhart gave a polynomial-time algorithm for Hamilton problem in solid grid graphs **[4]**. However, the algorithm cannot work without the assumption of a solid grid **[5]**. It remains open for whether a polynomial-time algorithm exists for solid grid graphs with some holes.

The study of cycle bases in a graph is beginning from MacLane's research on the characterization of planar graphs in 1937 **[6]**. Associated with the graph there is a vector space over GF(2), called the cycle space, consisting of the edge incidence vectors of all cycles (including the null cycle) and of all unions of edge-disjoint cycles of the graph. A set of cycles in the graph is a cycle basis if it is a basis in the cycle space of the graph. Any cycle in the graph can be written as a symmetric difference of the cycles in the basis. In this paper we survey the cyclic structure of Hamiltoncity in grid graphs with the cycle basis in which the cycles satisfy the condition of the tiling of the plane with polygons. Therefore, a Hamilton cycle can be denoted by a symmetric difference of the cycles in the subset of the cycle basis.

From the combinatorial relations of vertices in the subset, we deduce the formula for inside faces (the terms faces, polygons and cycles are used interchangeably in this paper) in Grinberg theorem **[7]**, called Grinberg equation in this paper.

Grinberg theorem is a necessary condition for a plane graph to be Hamiltonian. There have many works on its application in study of Hamilton graphs. 2003, A.N.M. Salman, E.T. Baskoro, and H.J. Broersma **[8]** characterizes the Hamiltoncity of the rectangular grid graphs and those included even holes. 2011, G.L. Chia and Carsten Thomassen **[9]** gave a short unified proof of Hamiltoncity for Petersen graphs using Grinberg theorem. However, there have no results in applications of polygonal grid graphs. By the deduction of Grinberg equation, we give a kind of cycles in polygonal grid graphs whose existence means that the graph is non-Hamiltonian, called non-Hamiltonian holes. Therefore, we characterize the existence condition of non-Hamiltonian holes in polygonal grid graphs and obtain the necessary and sufficient condition of a polygonal grid graph to be Hamiltonian. The result in this paper provides a new starting point for developing a polynomial-time algorithm for Hamilton problem in general grid graphs.

**2. Some Terms and Notations**

A graph is a polygonal grid graph if all cycles in basis of the graph consist of polygons tiling on a plane. We use G to denote a polygonal grid graph. The weight of edge $e$ denoted by $w(e)$ the sum of the number of cycles passing through $e$. Let $C_v$ be a cycle on vertex $v$ ($v \in V(G)$). $v$ is a boundary vertex if all its adjacent edges are $w \leq 2$ and there has $|C_v|-1$ incident edges of $w=2$. Edge $e$ is boundary if $w=1$. $v$ is a interior vertex if all its adjacent edges are $w=2$. See Figure 2.1.

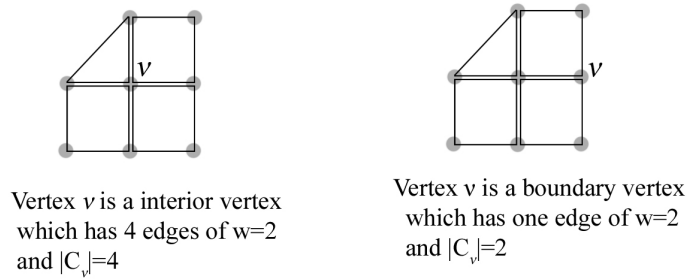

Vertex $v$ is a interior vertex which has 4 edges of w=2 and $|C_v|=4$

Vertex $v$ is a boundary vertex which has one edge of w=2 and $|C_v|=2$

Figure 2.1

A cycle is removable if no changes in the graph order when deleting it from G. Let E be a set of incident edges of vertex $p$. We use $d_2$ to denote an edge in E that another endvertex is of degree 2. There exist two cases on vertex $p$. (i)For $|E| \geq 3$ and $|d_2|=2$, it is clear that all other edges in E are not Hamilton path. (ii)For $|E| \geq 3$ and $|d_2| \geq 3$, the graph included E is not Hamiltonian. See Figure 2.2. We say G is claw($d_2$)-free if there have no these cases in graph G. We restrict the graphs to be claw($d_2$)-free polygonal grid graphs in the following.

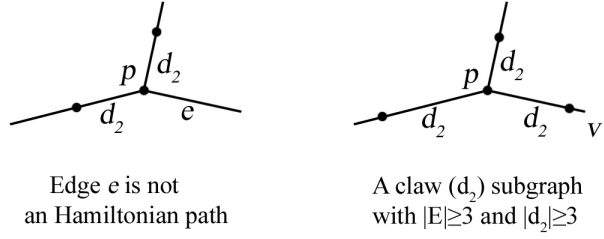

Edge *e* is not an Hamiltonian path    A claw ($d_2$) subgraph with $|E|\geq 3$ and $|d_2|\geq 3$

Figure 2.2

## 3. Grinberg Equation

First, we give Grinberg theorem as following:

**Theorem 3.1** *Suppose a plane graph has a Hamilton cycle C, such that there are $f_i'$ faces of length i inside C and $f_i''$ faces of length i outside C. then $\sum_i (i-2)(f_i'-f_i'')=0$.*

The formula $\sum (i-2)(f_i'-f_i'')=0$ is consist of $\sum(if_i'-2f_i')=|C|-2$ and $\sum(if_i''-2f_i'')=|C|-2$ **[10]**. $|C|-2$ denotes the length of C minus two. $\sum(if_i'-2f_i')=|C|-2$ is the formula of inside faces, called Grinberg equation in this paper, briefly the equation.

Next, we give the deduction of Grinberg equation from the relations of vertex sets in cycle basis of a graph. Let G be a claw($d_2$)-free polygonal grid graph, $B_p$ be a cycle basis of satisfied the condition of the tiling of the plane with polygons, $f$ denote a subset of $B_p$, $f_i$ be a cycle of order $i$ in $f$, and $V_i \in V(f_i)$. Known G can be expressed by all cycles in the $B_p$ and $V_i \in V(f_i)$, therefore the vertex set of G is equal to the union of all subsets $V_i$ in G. By inclusion and exclusion principle, the relation of the vertex set of G and $V_i$ can be expressed as the following

$$|\cup_{i=3}^{n} V_i| = \sum_{i=3}^{n} |V_i| - \sum_{3\leq i<j\leq n} |V_i \cap V_j| + \sum_{3\leq i<j<k\leq n} |V_i \cap V_j \cap V_k| - \ldots + (-1)^{n-1}|V_i \cap V_j \cap \ldots \cap V_n|. \quad (3.1)$$

Here $n=|E|-|V|+1$. We write equality (6.1) as $|\cup V|=\sum|V_i|-\sum|V_i\cap V_j|+\ldots+(-1)^{n-1}\sum|V_i\cap V_j\cap\ldots\cap V_n|$ in the following for short. Let $\sum|V_i\cap V_j|_{s\neq 2}$ denote the sum of item $|V_i\cap V_j|\neq 2$, $\sum|V_i\cap V_j|_{s=2}$ denote the sum of item $|V_i\cap V_j|=2$, and $\beta$ denote $\sum|V_i\cap V_j|_{s\neq 2}+\ldots+(-1)^{n-1}\sum|V_i\cap V_j\cap\ldots\cap V_n|$. Then, we can write equality (3.1) as following:

$$|\cup V|=\sum|V_i|-\sum|V_i\cap V_j|_{s=2}+\beta. \quad (3.2)$$

We say the subset $f$ is Hamiltonian if the symmetric difference of all cycles in $f$ is a Hamilton cycle in graph G. Let $|f|$ be the number of cycles in the subset.

**Proposition 3.1** *There has $\beta=0$ if $f$ is Hamiltonian.*

*Proof.* For $|f|=1$, there has only one cycle in the basis of G, we have $|\cup V|=\sum|V_i|$ and $\sum|V\cap|(n\geq 2)=\varnothing$,

substituting two equalities in (3.2), hence we have $\beta=0$. For $|f|\geq 2$, since the symmetric difference of cycles in $f$ is a Hamilton cycle of G, then there must exist two cases: (i) $\sum|V_i \cap V_j|_{s\neq 2}=\emptyset$. Otherwise, there would have a pair of cycles with a common edge of which the symmetric difference is a claw ($d_2$) subgraph (which satisfies $|E|\geq 3$ and $|d_2|\geq 3$), thus, $f$ is not Hamiltonian, contradiction. (ii) $\sum|V\cap|$ ($n\geq 3$)$=\emptyset$. Otherwise, the symmetric difference of every three cycles with a common edge would be a subgraph satisfied $|E|\geq 3$ and $|d_2|\geq 3$, and hence $f$ is not Hamiltonian, contradiction. Therefore, substituting (i) and (ii) in (3.2), we derive $\beta=0$. □

By the Proposition 3.1, we can write equality (3.2) in short as following:

$$|\cup V|=\sum|V_i|-\sum|V_i \cap V_j|_{s=2} \tag{3.3}$$

And,
(a) if $f$ is Hamiltonian, then the symmetric difference of all cycles in $f$ is a Hamilton cycle. This means that $\sum|V_i \cap V_j|_{s=2}=2(|f|-1)$. Since $|f|=|f_3|+|f_4|+\ldots+|f_n|$, then

$$\sum|V_i \cap V_j|_{s=2}=2(|f_3|+|f_4|+\ldots+|f_n|-1) \tag{3.4}$$

(b) Note that $\sum|V_i|$ is the sum of orders of all cycles in $f$, therefore,

$$\sum|V_i|=\sum|V_3|+\sum|V_4|+\ldots+\sum|V_n|. \tag{3.5}$$

Let $\sum|V_i|=3f_3$, $\sum|V_4|=4f_4$, ..., $\sum|V_n|=nf_n$. Then we can write Equality (3.5) as

$$\sum|V_i|=3f_3+4f_4+\ldots+nf_n. \tag{3.6}$$

(c) Known that $|V|=|\cup V_i|$, substituting equality (3.4), equality (3.6), and $|V|=|\cup V_i|$ in equality (3.3), we obtain

$$|V|=3f_3+4f_4+\ldots+nf_n-2(|f_3|+|f_4|+\ldots+|f_n|-1). \tag{3.7}$$

Hence,
$$\sum if_i-2(\sum f_i-1)=V. \tag{3.8}$$

After identical transformation, we can write equality (3.8) as $\sum if_i-2\sum f_i+2=|V|$, then we have $\sum(if_i'-2f_i')=|V|-2$, since here $|V|=|C|$ and $f_i$ is an interior face, then $\sum(if_i'-2f_i')=|C|-2$, which is the formula of inside faces in Grinberg theorem.

We call the equality (3.8) Grinberg equation, briefly in the equation. For a given basis of G, the cycles are called solution cycles if they can be inside faces satisfied the equation, and non-solution cycles if outside faces. We say the equation of G has solution if there has a partition of solution cycles and non-solution cycles.

## 4. Non-Hamiltonian Holes

In deduction of Grinberg equation, since the symmetric difference of all cycles in $f$ is a Hamilton cycle, then all pair of cycles with a common edge satisfy $\sum |V_i \cap V_j|_{s \neq 2} = \emptyset$. See the left graph in Figure 4.1. Clearly, it is Hamiltonian. While in the point of view of connection of two subsets of vertices, every pair of cycles without common edges also satisfy $\sum |V_i \cap V_j|_{s \neq 2} = \emptyset$. See the right graph in Figure 4.1. There exists a claw (d2) subgraph satisfied $|E| \geq 3$ and $|d_2| \geq 3$, so the graph is non-Hamiltonian. Hence, it is insufficient that using the equation of a graph has solutions to determine whether it is Hamiltonian.

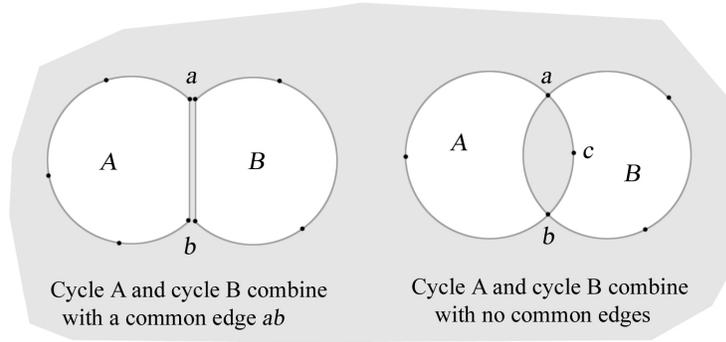

Cycle A and cycle B combine with a common edge $ab$  
Cycle A and cycle B combine with no common edges

Figure 4.1

When investigating the right graph with the cycle basis, the shadow region $abc$ ought to be a cycle in the basis. Since the set of vertices in region $abc$ equal to the intersection of the set of vertices in cycle A and the set of vertices in cycle B, then their intersection satisfy $\sum |V_i \cap V_j|_{s \neq 2} = \emptyset$ after removing the region $abc$ as a cycle in the basis. Thus, the equation of the right graph in Figure 4.1 has a solution. Known that cycle A and cycle B are not removable, hence the region $abc$ is the unique non-solution cycle to satisfy the solution of the equation of this non-Hamilton graph (the right graph in Figure 4.1), while the cycles with common edge $ab$ are solution cycles.

With the character of the shadow region $abc$, we give the following definitions.
Let G be a claw($d_2$)-free polygonal grid graph. Without confusion, we also use G to denote a cycle basis satisfied the condition of the tiling of the plane with polygons.

Let $x$ be a vertex of $deg(x) \geq 4$ in G, $C_x$ be a set of removable cycles on vertex $x$ such that $x$ can be turned into a boundary vertex of $deg(x) = 4$ by deleting $C_x$ from G which write as G−$C_x$ and satisfying G−$C_x$ has solutions. $C_k$ is a removable cycle including interior vertices but no boundary edges (edges of $w=1$) on vertex $x$ in G−$C_x$. $C_e$ is a set of the removable cycles jointing $C_k$ with common edges. $C_v$ is a set of cycles jointing $C_k$ with common vertices. $C_{xe}$ is a set of the removable cycles in G−$C_x$ jointing $C_k$ with the common vertex $x$.

On vertex $x \in V(G)$ in the given graph G having solutions of the equation, $C_k$ is a globally non-Hamiltonian hole if $C_k$ being a non-solution cycle is a necessary condition for the equation of G−$C_x$ ($C_x \neq \emptyset$) having an solution. See Figure 4.2(ii). For graph G having solutions of the equation of G−$C_x$−$C_k$, $C_k+C_e+C_v$ is a subgraph of G with $C_x=\emptyset$, $C_k$ is a locally non-Hamiltonian hole if there has a set of cycles in $C_k+C_e+C_v$ in which $C_k$ is a globally non-Hamiltonian hole. See Figure 4.2(i).

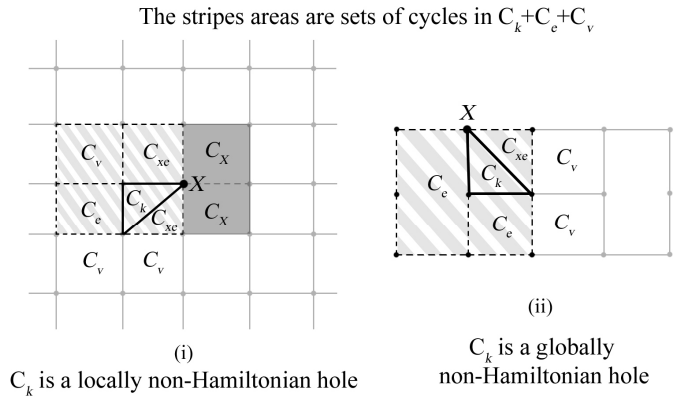

The stripes areas are sets of cycles in $C_k+C_e+C_v$

(i)
$C_k$ is a locally non-Hamiltonian hole

(ii)
$C_k$ is a globally non-Hamiltonian hole

Figure 4.2

Given graph G having solutions of the equation, $(G-C_x-C_{xe})-\Sigma(C_x+C_{xe}+C_k)$ denote a subgraph obtained by removing $C_x$ and $C_{xe}$ on a beginning vertex $x$ and then continuously removing $C_x$, $C_{xe}$, $C_k$ from the produced subgraph having solutions of the equation until no $C_k$ in it. For example in Figure 4.3, beginning with $x_0$ and removing $C_{xe0}$, we obtain a produced subgraph $G-C_{x0}-C_{xe0}$ having solutions of the equation, then we select $x_1$ randomly as a beginning vertex and remove $C_{xe1}$ and $C_{k1}$ from G, repeating this procedure until we get to the last beginning vertex $x_{10}$ and remove $C_{xe10}$ and $C_{k10}$ from $(G-C_{x0}-C_{xe0})-\Sigma(C_{x9}+C_{xe9}+C_{k9})$, and then we obtain a produced subgraph $(G-C_{x0}-C_{xe0})-\Sigma(C_{x10}+C_{xe10}+C_{k10})$.

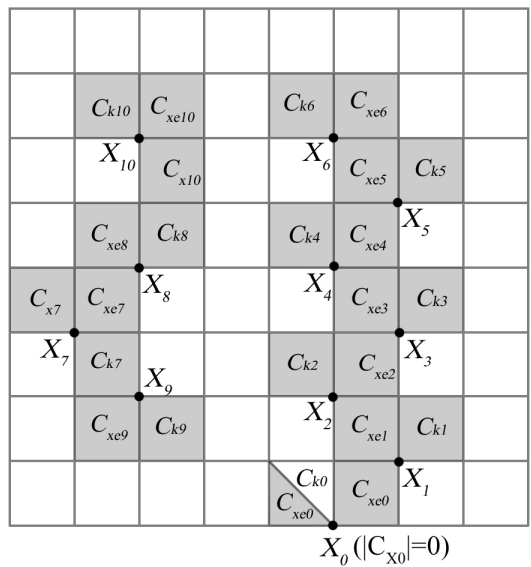

Begining with $X_0$ we have $(G-C_{x0}-C_{xe0})-\Sigma(C_x+C_{xe}+C_k)$
which is a subgraph of G

Figure 4.3

**Lemma 4.1** *A $C_k$-free graph G is Hamiltonian if and only if there has solutions of the equation of G.*

*Proof.* Let G be a $C_k$-free and claw($d_2$)-free polygonal grid graph. By Theorem 3.1, If G has no solutions of the equation, then G is non-Hamiltonian no matter there has $C_k$ or not. If G has solutions of the equation, then it means there is a partition of solution cycles and non-solution cycles in the cycle basis of G. Known the symmetric difference of the set of solution cycles is a spanning subgraph of G (the set of solution cycles is equal to a subset $f$ of $B_p$ in section 3), which satisfies $\sum|V_i \cap V_j|_{s \neq 2} = \varnothing$ and $\sum|V \cap| (n \geq 3) = \varnothing$ for every two incident cycles ( each of these cycles can be a cycle produced from the symmetric difference of cycles in the spanning subgraph, see example in Figure 4.4(b), $C'_2 = C_2 \oplus C_5$) in it, that is $\beta = 0$. Thus, there exists $\sum|V_i \cap V_j|_{s=2}$ only for those every two cycles. While, there have two cases we must consider: two incident cycles combined with a common edge or without a common edge. Note that G is a $C_k$-free graph, every two incident cycles in a spanning subgraph of G is combined with a common edge. See the only two types of these combination in Figure 4.4: $\{C_1, C_2, C_3, C_4\}$ and $\{C_1, C_2, C_3, C_4, C_5\}$, and the intersections of two set of cycles are one cycle, a spanning cycle in graph G. Hence, G is Hamiltonian. □

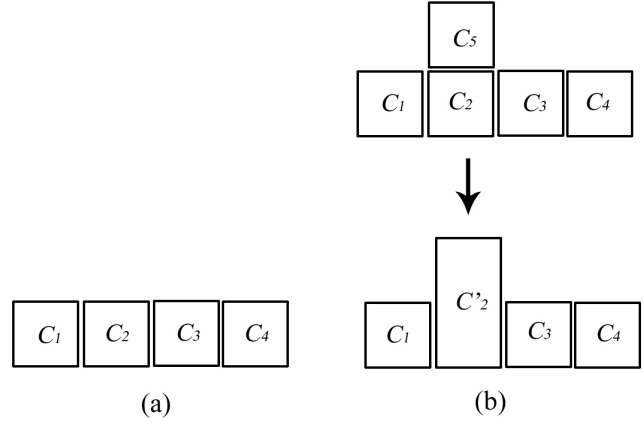

Figure 4.4  Two cases of the cycle sets satisfied $\beta=0$

**Lemma 4.2** *Given graph $G-C_x$ having solutions of the equation, $C_x$ is a locally non-Hamiltonian hole of $G-C_x$. $C_k$ is a globally non-Hamiltonian hole if and only if there is no solutions of the equation of $(G-C_x-C_{xe})-\Sigma(C_x+C_{xe}+C_k)$*

*Proof.* Given graph $G-C_x$ having solutions of the equation.

For not having the solutions of the equation of $(G-C_x-C_{xe})-\Sigma(C_x+C_{xe}+C_k)$, by definition, under the condition of that having solutions of the equation of $G-C_x-C_{xe}$, $(G-C_x-C_{xe})-\Sigma(C_x+C_{xe}+C_k)$ is a $C_k$-free graph, by Lemma 4.1, $(G-C_x-C_{xe})-\Sigma(C_x+C_{xe}+C_k)$ is non-Hamiltonian. This implied $C_k$ can not be a solution cycle. Note $C_x$ is a locally non-Hamiltonian hole of $G-C_x$, thus $C_k$ can only be a non-solution cycle of G under the condition of that having solutions of the equation of $G-C_x-C_k$, and known $G-C_x$ is a graph having solutions of the equation, and hence $C_k$ is a globally non-Hamiltonian hole of G. For having the solutions of the equation of $(G-C_x-C_{xe})-\Sigma(C_x+C_{xe}+C_k)$, similarly, $(G-C_x-C_{xe})-\Sigma(C_x+C_{xe}+C_k)$ is Hamiltonian. It means $C_k$ can be a solution cycle of G. Then $C_k$ is not a globally non-Hamiltonian hole of G. □

**Lemma 4.3** *G is non-Hamiltonian if $C_k$ is a globally non-Hamiltonian hole of G.*

*Proof.*    By definition we know that $C_k$ being a non-solution cycle is a necessary condition for the equation of $G-C_x$ ($C_x \neq \emptyset$) having an solution if $C_k$ is a globally non-Hamiltonian hole of G. Then we have that there has a solution of the equation of $G-C_x-C_k$, and there must be two cycles passed through vertex *x*, which is contradiction to the definition of Hamilton graph. Thus, G is non-Hamiltonian.    □

## 5. Conclusion

**Theorem 5.1** *Let G denote a no globally non-Hamiltonian hole and claw($d_2$)-free polygonal grid graph. G is Hamiltonian if and only if the equation of G has solutions.*

*Proof.*    If the equation of G has not solutions then it is a non-Hamiltonian by theorem 3.1. Consider the equation of G has solutions, know G is no globally non-Hamiltonian hole, for having $C_k$ in the graph, by Lemma 4.2, there has solutions of the equation of $(G-C_x-C_{xe})-\Sigma(C_x+C_{xe}+C_k)$, and then by Lemma 4.1, $(G-C_x-C_{xe})-\Sigma(C_x+C_{xe}+C_k)$ is Hamiltonian. Clearly G is Hamiltonian. For having no $C_k$ in the graph, by Lemma 4.1 we directly obtain G is Hamiltonian.    □

## 6. Independent Subbases

From Theorem 5.1, for a Hamilton graph G, we know that there is a set of cycles in the cycle basis whose symmetric difference is a Hamilton cycle of G. Every cycle in the set as a member has following property: the whole set is non-Hamiltonian if anyone in the set is non-Hamiltonian. Without loss of generality, we extend these members to subgraphs.

In the basis of a polygonal grid graph, the boundary-element set is a set of cycles including boundary vertices or boundary edges. Independent Subbasis is the interior connected subgraph bounded by a subset of boundary-element cycles. Therefore, a polygonal grid graph is consisted of boundary-element cycles and the independent Subbases they bounded. The minimal boundary-element set is a set of boundary-element cycles such that it can not bound any interior connected subgraph when removing any cycle from the set. The boundary-element co-set is the set of cycles having no relations with minimal boundary-element sets. We use *g* to denote a subgraph consisting of the independent Subbases and the minimal boundary-element set. |*g*| is the number of *g* in the graph. By the definitions above, it is clearly that all the graphs we consider before is of the basis of |*g*|=1.

Obviously, subgraph *g* is equal to a cycle, denoted by $C_g$, with the same order of *g* if it is Hamiltonian. Let $G_g$ be a graph derived from the union of $C_g$ and the boundary-element co-set in graph G. Then we have the following proposition directly:

**Proposition 6.1** *The bases of G and $G_g$ have the same Hamiltoncity.*

By Proposition 6.1, we can use $G_g$ to study the Hamiltoncity of a basis of |*g*|≥2. See the left graph in Figure 6.1, it has two independent Subbases (in spite of that they are empty graph) and one boundary-element co-set (cycle C). Both independent Subbases is Hamiltonian, thus the bases of G can

be expressed by the bases of $G_g$ which is the union of two $C_9$ and one $C_4$. See the right graph in Figure 6.1.

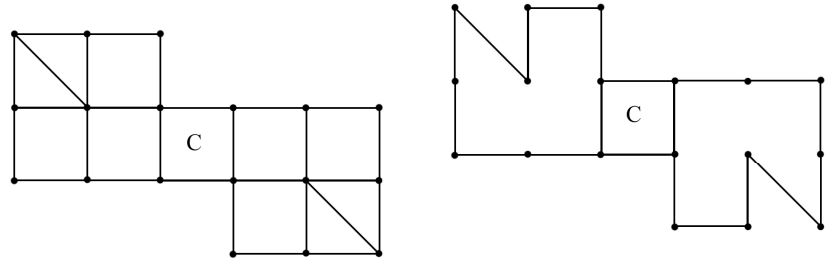

Figure 6.1

In the following we take Tutte graph in Figure 6.2 as another example. Tutte graph is a no globally non-Hamiltonian hole and claw($d_2$)-free polygonal grid graph whose basis is of $|g|=3$. There are three independent Subbases according to the parts of heavy lines in the graph. Each equation of $G_g$ is $10f_{10}+5f_5+4f_4-2(f_{10}+f_5+f_4-1)=25$, there has a solution for the equation of $G_g$. By theorem 5.1, each $G_g$ is Hamiltonian, and then we can substitute $C_{25}$ to $G_g$. And the boundary-element co-set is empty. Hence, the equation of Tutte graph is $25f_{25}-2(f_{25}-1)=46$.

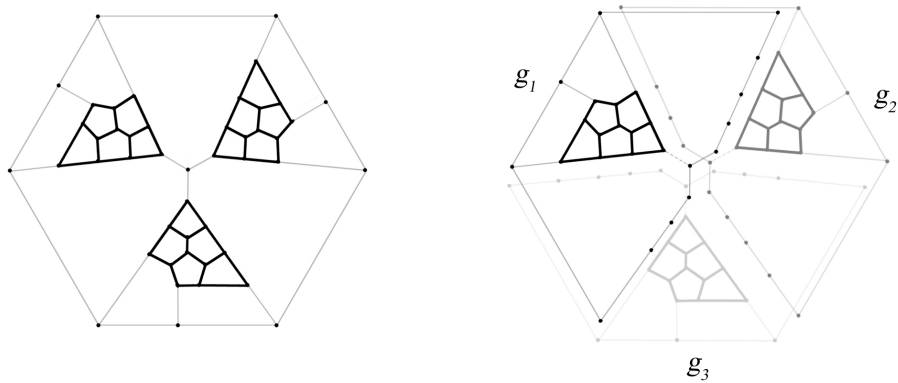

Figure 6.2  Tutte graph consists of three $C_{25}$ lapped over each other

**References**
[1] Reinhard Diestel. Graph theory. Second Edition, Springer-Verlag New York, Lnc., 2000.
[2] Alon Itai, Christos H. Papadimitriou and Jayme Luiz Szwarcfiter, Hamilton Path in Grid Graphs. SIAM J. Comput. Vol. 11, No. 4, November 1982.
[3] Hwan-Gue, Cho and Alexander Zelikovsky, Spanning Closed Trail and Hamiltonian Cycle in Grid Graphs. In Proceedings of the 6[th] international symposium on Algorithms and Computations (ISAAC), Australia, page 342. Springer Verlag, 1995.
[4] C.Umans and W.Lenhart. Hamiltonian Cycles in Solid Grid Graphs. In Proceedings of the 38[th] Symposium on Foundations of Computer Science, pages 496-505. IEEE, 1997.

[5] Lazaros Koromilas. Grid Hamiltonicity. May 27, 2011.
   http://www.csd.uoc.gr/~hy583/2010_presentations/Grid_Hamiltonicity.pdf

[6] Saunders Mavlane (Cambridge, Mass.). A Combinatorial Condition for Planar Graphs. Fundamenta Mathematicaae 28, 22-32. 1937.

[7] Grinberg E.J., Plane homogeneous graphs of degree three without Hamiltonian circuits. Latvian Math. Yearbook 5 (1968), 51~58.

[8] A.N.M. Salman, E.T. Baskoro, and H.J. Broersma. A Note Concerning Hamilton Cycles in Some Classes of Grid Graphs. PROC. ITB Sains & Tek. Vol. 35 A, 1, 2003, 65-70.

[9] G.L. Chia and Carsten Thomassen. Grinberg's Criterion Applied to Some Non-Planar Graphs. Ars Combinatoria 100(2011) 3-7.

[10] Douglas B. West, Introduction to graph theory, Seconded, Pearson Education Asia Ltd., 2004, 302.